\documentclass[11pt,showpacs,preprintnumbers,amsmath,aps,amssymb]{revtex4}
\usepackage{graphicx}
\usepackage{dcolumn}
\usepackage{bm}

\begin{document}
\def\sqr#1#2{{\vcenter{\vbox{\hrule height.#2pt
        \hbox{\vrule width.#2pt height#1pt \kern#1pt
           \vrule width.#2pt}
        \hrule height.#2pt}}}}
\def\square{\mathchoice\sqr54\sqr54\sqr{6.1}3\sqr{1.5}6}
\font\tm=cmss10 scaled 750
\title{Inflation from Supersymmetric Quantum Cosmology}
\author{J. Socorro$^1$}
\email{socorro@fisica.ugto.mx}
\author{Marco D'Oleire$^2$}
\email{marcodoleire@gmail.com}

\affiliation{$^1$Departamento de  F\'{\i}sica, DCeI, Universidad de Guanajuato-Campus Le\'on,\\
A.P. E-143, C.P. 37150, Le\'on, Guanajuato, M\'exico\\
$^2$Facultad de Ciencias de la Universidad Aut\'onoma del Estado de
M\'exico, \\
Instituto Literario  No. 100, Toluca, C.P. 50000, Edo de Mex, M\'exico \\
}%

\date{\today}

\begin{abstract}
We derive a special scalar field potential using the
anisotropic Bianchi type I cosmological model from canonical quantum cosmology under
determined conditions in the evolution to anisotropic variables $\beta_\pm$. In the process, we obtain a 
 family of potentials that has been introduced by hand in the literature to explain  cosmological data. 
 Considering supersymmetric quantum cosmology, this family is scanned, fixing the exponential potential as more
 viable in the inflation scenario $\rm V (\phi) = V_0 \,e^{-\sqrt{3}\phi}$.
\end{abstract}

\maketitle

\section{Introduction}

One of the main problems of inflationary cosmology is to find a mechanism to derive in a natural way 
the appropriate scalar field potential in order to develop  enough e-foldings of inflation. 
By natural, we understand a mechanism from which some theory provides a scalar field potential that offers 
the convenient features of inflation.
In this work we derive a scalar field potential from supersymmetric 
quantum cosmology that gives these conditions.

In a previous work, we determined  scalar potentials from an exact solution to 
the Wheeler-DeWitt (WDW) equation in the quantum cosmology scenario \cite{wssa}, using as a toy model an homogenous 
and isotropic cosmological model.  There we focus on solutions that may be relevant for the early universe 
constructed within of WKB approximation. 
Recently, these scalars potentials were obtained using a
local supersymmetric scheme \cite{celia}. Nowadays it is a common issue in cosmology to make use 
of scalar fields $\phi$ as the responsible agents of some of the most intriguing aspects of 
our universe \cite{scalar-1, scalar-2, scalar-3, sca4, sca5, sca6, sca7, sca8, sca9}, such as inflation 
\cite{copeland1, tsuji}, dark matter  and dark energy \cite{copeland}.
 The natural derivation of a scalar potential is a challenge, posing the following question: What
 physical processes provide the adecuate scalar field potentials that govern the universe in determined epoch? 
 To answer  this question, 
we use the ideas of quantum cosmology to solve the Wheeler-DeWitt equation with a  particular
ansatz for the Bianchi type I universe wave-function. In this scheme, we obtain two possible
scenarios, the first one with an  scalar exponential potential  $\rm V(\phi)=V_0\, e^{\lambda \phi}$,
 and  the second one giving a family of potentials, similar to those obtained in our previous work \cite{wssa}. 
It is interesting that in the first scenario the $\lambda$ parameter is not fixed by the quantum scheme, remaining  
as a free parameter of the theory. To fix it we invoke supersymmetric scale, 
using the tools of supersymmetric quantum cosmology in order to find most viable 
scalar potential  for the inflationary epoch, in this scale. 
To do this, we applied supersymmetry as a square root of general relativity \cite{bene, osb, lidsey, socorro}, 
in which the Grassmann variables are only auxiliary and can not 
identified as the supersymmetric partners of the cosmological bosonic variables.
Therefore, we  construct a family of scalar potentials treating the quantum solutions 
to anisotropic Bianchi type I cosmological model in the anisotropic variables $\beta_+$ and $\beta_-$. 
The conditions we use give us a special structure for the scalar potential; 
 By simplifying the Wheeler-DeWitt equation we obtain two cases:  one in which both parameters  $\beta_\pm$
 have  hyperbolic trigonometric functions as solutions, and another where
$\beta_-$ ($\beta_+$)  have a trigonometric  (hyperbolic trigonometric)  behavior. 
This potential is also a good candidate, depending on the parameter value, in order to study
inflation, dark matter, dark energy or  tachyon  models \cite{taquion}. The transform Wheeler-DeWitt equation 
can be solved using a particular ansatz in the WKB approximation (Bohmian representation \cite{bohm}). This
method has been used in the literature \cite{OS} to solve the cosmological Bianchi class A models, 
and in a particular, our result in the second case is similar to the one found in reference \cite{wssa} 
for the isotropic Friedmann-Robertson-Walker (FRW) cosmological model. On the other hand, 
the best candidates quantum solutions become these that have a damping behavior 
with respect to the scale factor, in sense that we obtain a good classical solution using the 
WKB approximation in any scenario in the evolution of our universe \cite{HaHa,hawking}. The supersymmetric
scheme have the particularity that is very restrictive  because there is more constraints
equations applied to the wavefunction. So, in this work we found that exist a tendency for supersymmetric vacua
to remain close to their semiclassical limits, because the exact solutions found are also the lowest-order WKB approximation,
and not corresponds to the full quantum solutions found previously.

\section{The Wheeler-DeWitt equation}
On the Wheeler-DeWitt equation there are a lot of papers dealing with different problems, for example,
Gibbons and Grishchuk \cite{gibbons} asked the question of what a typical wave function for the universe is. In reference
\cite{ruffini} appears one excellent recopilation of paper on quantum cosmology where
the problem of how the universe emerged from Big Bang singularity can no longer be neglected. Also, an important
approach to this problem is the wave function proposal in which the universe would be completely self-contained
without any singularities and without any adges.
Our goal in this paper deals with the problem to built the appropriate scalar potential in the inflationary scenario.

We start by recalling the canonical formulation of the ADM formalism to the
diagonal Bianchi Class A cosmological models. The metrics have the form
\begin{equation}
\rm ds^2= -(N^2- N^j N_j)dt^2 + e^{2\Omega(t)} e^{2\beta_{ij}(t)} \,
\omega^i \omega^j,
\label{metrica}
\end{equation}
where $\rm N$ and $\rm N_i$ are the lapse and shift functions
respectively, $\Omega(t)$ is a scalar and $\rm \beta_{ij}(t)$ a 3x3
diagonal matrix, $\rm \beta_{ij}=diag(\beta_+ +\sqrt{3}
\beta_-,\beta_+ -\sqrt{3} \beta_-, -2\beta_+)$, $\rm \omega^i$ are
one-forms that  characterize  each cosmological Bianchi type model,
and that obey $\rm d\omega^i= \frac{1}{2} C^i_{jk} \omega^j \wedge
\omega^k,$ $\rm C^i_{jk}$ the structure constants of the
corresponding invariance group \cite{ryan}. The metric for the
Bianchi type I, takes the form
\begin{equation}
\rm ds^2_I=  - N^2dt^2 + e^{2\Omega} e^{2\beta_+ +2\sqrt{3}\beta_-}
dx^2 +e^{2\Omega} e^{2\beta_+ -2\sqrt{3}\beta_-} dy^2 + e^{2\Omega}
e^{-4\beta_+ } dz^2, \label{bi}
\end{equation}
the total lagrangian density function is given by
\begin{equation}
\rm {\cal L}_{Total}={\cal L}_g + {\cal L}_{\Lambda} +{\cal L}_{matter,\phi} =\sqrt{-g}\, (R-2\Lambda)+ {\cal L}_{matter,\phi},
\label{lagraq}
\end{equation}
we use as a first approximation a perfect fluid and a scalar field
as the matter content,  in a comoving frame \cite{ryan},
\begin{equation}
\rm {\cal L}_{Total}=\sqrt{-g}\, (R-2\Lambda+ 16\pi G \rho +
\frac{1}{2} g^{\mu\nu}\partial_\mu \phi \partial_\nu \phi + V(\phi) ),
\label{lagra}
\end{equation}
and  using (\ref{bi}) we have
\begin{equation}
{\cal L}_{Total} = e^{3\Omega}\left[\frac{6}{N}\dot{\Omega}^2-\frac{6}{N}\dot{\beta_+}^2-\frac{6}{N}\dot{\beta_-}^2
-\frac{6}{N}\dot{ \varphi}^2
+16\pi GN\rho-2\Lambda N+\frac{V(\varphi)N}{2}\right],\label{lagra1}
\end{equation}
where we redefined the original scalar field as $\rm \phi=\sqrt{12}  \varphi$.

The corresponding momentas are calculated in the usual way
\begin{eqnarray*}\rm
\Pi_\Omega &=& \rm \frac{\partial{\cal L}}{\partial\dot{\Omega}}=\frac{12\dot{\Omega}}{N}e^{3\Omega} 
\rightarrow  \dot{\Omega}= \frac{e^{-3\Omega}}{12}N\Pi_\Omega\\
\rm \Pi_+ &=&\rm \frac{\partial{\cal L}}{\partial\dot{\beta_+}}=-\frac{12\dot{\beta_+}}{N}e^{3\Omega}  
\rightarrow \dot{\beta_+}=-\frac{e^{-3\Omega}}{12}N\Pi_+\\
\rm \Pi_- &=&\rm \frac{\partial{\cal L}}{\partial\dot{\beta-}}=-\frac{12\dot{\beta_-}}{N}e^{3\Omega} 
\rightarrow  \dot{\beta_-}=-\frac{e^{-3\Omega}}{12}N\Pi_-\\
\rm \Pi_\varphi &=&\rm  \frac{\partial{\cal L}}{\partial\dot{\phi}}=-\frac{12\dot{\varphi}}{N}e^{3\Omega}  
\rightarrow  \dot{\varphi}=-\frac{e^{-3\Omega}}{12}N\Pi_{\varphi}\\
\end{eqnarray*}
now writing (\ref{lagra1}) in canonical form  $\rm {\cal L}_{can}=\Pi_q\dot{q}-N{\cal H}$ where $\rm {\cal H}$ is the
hamiltonian density function, 
\begin{eqnarray*}
{\cal L}_{can}&=&\Pi_\Omega\dot{\Omega}+\Pi_+\dot{\beta_+}+\Pi_-\dot{\beta_-}+\Pi_\varphi\dot{\varphi}\\
&&-\frac{Ne^{-3\Omega}}{24} \left( \Pi_\Omega^2-\Pi_+^2-\Pi_-^2-\Pi_\varphi^2
-e^{6\Omega}\left[384\pi G\rho-48\Lambda+12V(\varphi)\right]  \right)
\end{eqnarray*}
we obtain the corresponding Hamiltonian density
function
\begin{eqnarray}
\rm {\cal H} &=& \frac{e^{-3\Omega}}{24}\left ( \Pi_\Omega^2-\Pi_+^2-\Pi_-^2-\Pi_\varphi^2
-e^{6\Omega}\left[384\pi G\rho-48\Lambda+12V(\varphi)\right]\right )
\end{eqnarray}  \label{Hamiltonn}
when we include the energy-momentum tensor for a barotropic perfect fluid $\rm p=\gamma \rho$, we have
\begin{equation}
\rm {\cal H}= \frac{e^{-3\Omega}}{24} \left(\Pi_\Omega^2 - \Pi_+^2 -\Pi_-^2 -
\Pi_\varphi^2 +48\Lambda  e^{6\Omega} -384\pi G M_\gamma e^{-3(\gamma-1)\Omega} -
12e^{6\Omega}V(\varphi) \right). \label{hami1}
\end{equation}

Imposing the quantization condition and applying this hamiltonian to the wave  function $\Psi$, we obtain 
the WDW equation for these models  in the minisuperspace  by the usual identification
$\rm P_{q^\mu}$
by $\rm -i \partial_{q^\mu}$ in (\ref {hami1}),
with $\rm q^\mu=(\Omega, \beta_+,\beta_-,\varphi)$, and
following Hartle and Hawking {\cite {HaHa}} we consider a
semi-general factor ordering which  gives
\begin{equation*}
\hat{H}\Psi=\left[-\frac{\partial^2}{\partial\Omega^2}+\frac{\partial^2}{\partial\beta_+^2}+
\frac{\partial^2}{\partial\beta_-^2}+\frac{\partial^2}{\partial\varphi^2}+Q\frac{\partial}{\partial\Omega}
+48\Lambda  e^{6\Omega} -384\pi G M_\gamma e^{-3(\gamma-1)\Omega} -
12e^{6\Omega}V(\varphi)\right ]\Psi=0,
\end{equation*}
where Q measures the ambiguity in the factor ordering between the scalar function $\Omega$ and its corresponding momenta. 
This equation is not easy to solve,  first because we do not have the
structure of scalar potential, and second, it depends strongly on the class of scenario we analyze
with barotropic equation. In the follow, for simplicity we shall use the inflationary case, $\gamma=-1$

Using the following ansatz for the wavefunction $\Psi(\Omega,\varphi,\beta_\pm)=e^{\pm a_1 \beta_+ \pm a_2 \beta_-} \Xi(\Omega,\varphi)$,  
we obtain a reduced WDW
\begin{equation}
  \rm  \left[-\frac{\partial^2}{\partial \Omega^2} +\frac{\partial^2}{\partial \varphi^2}  +Q\frac{\partial}{\partial \Omega}
  +e^{6\Omega}\left(48\Lambda  - 384\pi G M_{-1} - 12  V(\varphi)\right) + c^2\right]\Xi=0 , \label{wdw2}
\end{equation}
where the constant $\rm  c^2=a_1^2+a_2^2$.
 
Eqn (\ref{wdw2}) can be written in compact form as
\begin{equation} 
\Box \, \Xi + Q \frac{\partial \Xi}{\partial \Omega} -U(\Omega,\varphi,\lambda_{eff}) \Xi =0 \, , 
\label {WDW}
\end{equation}
where the d'Alambertian in two dimensions is redefined as $\Box \equiv - \partial^2_\Omega +  \partial^2_\varphi$,
$\lambda_{eff}=48\Lambda- 384\pi G M_{-1} $ is the effective cosmological constant, 
and the potential $\rm U(\Omega,\varphi,\Lambda)=e^{6\Omega}\left[ 12  V(\varphi)-\lambda_{eff} \right] - c^2$.

To solve (\ref{wdw2}) we take the ansatz, which is similar to the one used in the Bohmian formalism into quantum mechanics \cite{bohm}
\begin{equation}
\Xi(\Omega, \varphi) = W(\Omega, \varphi) e^{- S(\Omega, \varphi)} \, ,
\label{wavefunction}
\end{equation}
where $S(\Omega, \varphi)$ is the {\it superpotential} function. Eq  (\ref{WDW}) 
can be written as the following set of partial differential equations
\begin{subequations}
\label{WDWa}
\begin{eqnarray}
(\nabla S)^2 - U &=& 0, \label{hj} \\
  W \left( \Box S + Q \frac{\partial S}{\partial \Omega}
  \right) + 2 \nabla \, W \cdot \nabla \, S &=& 0 \, , \label{wdwho} \\
  \Box \, W + Q \frac{\partial W}{\partial \Omega} & = & 0 \label{cons}  \, , 
\end{eqnarray}
where the first equation is the classical Hamilton-Jacobi equation, which plays an important role in this work. The different
terms in this equation are 
\begin{displaymath} \nabla W \cdot \nabla S \equiv - \left(
\partial_\Omega W \right) \left(\partial_\Omega S\right) + \left( \partial_\varphi W \right) \left(
\partial_\varphi S \right),  \quad 
(\nabla)^2 \equiv - \left(\partial_\Omega \right)^2 + \left( \partial_\varphi \right)^2.
\end{displaymath} 
\label{WDWecs}
\end{subequations}
Any exact solution complying with the set of equations (\ref{hj},\ref{wdwho}, \ref{cons}) will
also be an exact solution of the original WDW equation. Following reference \cite{wssa},
first we shall choose to solve eqns (\ref{hj}) and (\ref{wdwho}), whose solutions at the end will have to 
fulfill with eqn (\ref{cons}) which play the role of a constraint equation.

Taking the ansatz  \cite{kasper} \footnote{Similar ansatz was employed by Kasper in the WKB approximation of the
WDW equation in fourth-order quantum cosmology} 
\begin{equation}
\rm S(\Omega,\varphi)= \frac{1}{\mu}e^{3\Omega} g(\varphi) + c(b_1\Omega+b_2 \Delta \varphi),
\label{superpotential}
\end{equation}
with $\Delta \varphi=\varphi-\varphi_0$, $\varphi_0$ is a constant scalar field, 
$\rm b_i$ arbitrary constants, Eq (\ref{hj}) is transformed as
\begin{equation}
\left[-\frac{9}{\mu^2} g^2+\frac{1}{\mu^2}\left( \frac{dg}{d\varphi}\right)^2
-12V(\varphi)+\lambda_{eff}\right] e^{6\Omega}+ c^2\left[1-b_1^2+b_2^2\right]
+\frac{6c}{\mu}\left[\frac{b_2}{3} \frac{dg}{d\varphi}-b_1 g \right]e^{3\Omega}=0. \label{master}
\end{equation}
This equation is more dificult to solve, at this point we introduce the main idea of the paper to
obtain the  scalar potential family, which are strongly dependents to solutions for the
anisotropic variables $\rm \beta_\pm$. We include two steps to solve equation (\ref{master}):
\begin{enumerate}
\item{} First, consider that the second  and third parenthesis are null, but maintaining that 
$\rm c\not=0$. The first condition implies that $\rm b_1=\sqrt{1+b_2^2}$, and  the constants 
$\rm a_1$ and $\rm a_2$ are real, the solutions for the anisotropic variables $\beta_\pm$ can be considered
as hyperbolic trigonometric functions. The second condition becomes an ordinary differential equation
for  the unknown function $\rm g(\varphi)$, yielding  
\begin{equation}
\rm g(\varphi)=g_0 e^{\frac{\alpha}{2}\Delta \varphi }, \label{g0}
\end{equation}
with  $\rm g_0$ an integration constant and $\alpha=\frac{6b_1}{b_2}=\frac{\pm 6\sqrt{1+b_2^2}}{b_2}$ with $\rm b_2\not=0$. 
The scalar potential function become, when we
take  the first parenthesis in eqn (\ref{master}), 
\begin{equation}
\rm V(\varphi)= \left( 4 \Lambda- 32\pi G M_{-1}\right)+ V_0 e^{\alpha \Delta \varphi}, \label{potential}
\end{equation} 
with $\rm V_0=\frac{3g_0^2}{4 b_2^2\mu^2}$. With these results, the superpotential function 
(\ref{superpotential}) is
\begin{equation}
\rm S(\Omega,\varphi)= \frac{g_0}{\mu}e^{3\Omega} e^{\frac{\alpha}{2}\Delta \varphi} + c\left(\pm \sqrt{1+b_2^2}\Omega+b_2 \Delta \varphi \right).
\label{super}
\end{equation} 
Sustituting (\ref{super}) into (\ref{wdwho}), the corresponding solutions for the function W in the form $\rm e^{q\Omega + \eta \varphi}$ 
become $\rm W=Exp \left[ \frac{1}{2}(Q-3) \Omega -\frac{1}{4}\alpha \Delta \varphi  \right]$,
 we develop the following wavefunction
\begin{equation}
\rm \Psi=Exp\left[a_1 \beta_+ + a_2\beta_-  + \frac{1}{2}(Q-3) \Omega -\frac{1}{4}\alpha \Delta \varphi \right]\,\, 
e^{  -\frac{e^{3\Omega+\frac{\alpha}{2}\Delta \varphi}}{\mu}},
\label{wave1}
\end{equation}
with the constraint on the parameter $\alpha \leq \pm 6$.

In the literature \cite{cimento} the scalar potencial type $\rm V(\phi)=e^{\lambda \phi}$ gives a power law in the classical scale factor,
considering the flat FRW cosmological model when $\lambda < -\sqrt{2}$. 
If we consider the extreme values for the $\alpha$ parameter (Q=0) and the corresponding
transformation between $\varphi \to \phi$, we obtain the special scalar potential $\rm V(\phi)=V_0\, e^{\pm \sqrt{3} \phi}$,
 for standard inflationary model, this class of
potential has the advantage that classical analytical solutions can be found and for appropriate values
of the parameters, inflation can be obtained. 

\item{} In the second step, we consider that the constant c=0, implying that $\rm a_1=\pm i a_2$, then 
the  solutions for the wave function for the anisotropic variables $\beta_\pm$ are considered
trigonometric functions for the variable $\beta_+$ and hyperbolic trigonometric function for the variable $\beta_-$. Thus 
the superpotential term (\ref{superpotential}) has the simple form 
\begin{equation}
\rm S(\Omega,\varphi)= \frac{1}{\mu}e^{3\Omega} g(\varphi), \label{superpo}
\end{equation}
and equation (\ref{master}) becomes an ordinary differential equation for the unknown function
$\rm g(\varphi)$ in terms of the scalar potential $\rm {\cal V}(\varphi,\lambda_{eff})=V(\varphi)-\frac{\lambda_{eff}}{12}$,
\begin{equation}
\rm \left( \frac{dg}{d\varphi}\right)^2-9 g^2(\varphi)
=12\mu^2\left(V(\varphi)-\frac{\lambda_{eff}}{12}\right)=12\mu^2{\cal V}(\varphi,\lambda_{eff}), \label{de}
\end{equation}
this equation is similar to the one obtained in reference \cite{wssa}.
It is not surprising that this equation is similar to eqn (12) in reference \cite{wssa}, because
the anisotropic Bianchi type I cosmological model is the generalization of the flat FRW model. The last equation
has several exact solutions, which can be generated in the following way. Consider that ${\cal V}=g^2F(g)$,
where $\rm F(g)$ is an arbitrary function of its argument. So, eq. (\ref{de}) can be written
in cuadratures as
\begin{equation}
\rm \Delta \varphi=\pm \frac{1}{2\sqrt{3}}\int \frac{d ln g}{\sqrt{\frac{3}{4}+\mu^2 F(g)}}.
\end{equation}
In this way, we can solve  the $\rm g(\varphi)$ function, and then use the expression for the
potential term ${\cal V}=g^2F(g)$ back again to find the corresponding scalar potential that
leads to an exact solution to the Hamilton-Jacobi equation (\ref{hj}). Some examples are
shown in Table I.

\begin{center}
\begin{table}[h]
\caption{ \label{t:solutions} Some exact solutions to eq. (\ref{de}) and their corresponding
scalar potentials, where n is any real number and $V_0$ is an arbitrary constant, different to step one.
The third line is equivalent to this obtained by the first step.}
\begin{tabular}{|c|c|c|}
\hline
$\rm  F(g)$ & $\rm g(\varphi)$ &$\rm   {\cal V}(\varphi,\lambda_{eff})$ \\ 
\hline
$ 0 $ & $\rm Exp\left[\pm 3 \Delta \varphi\right]$& $ 0 $ \\
$ V_0 g^{-2} $ &$\rm {\sqrt{\frac{4\mu^2 V_0}{3}}} sinh(\pm 3 \Delta \varphi)$&  $ V_0 $ \\
$ V_0$ & $e^{\frac{\alpha}{2} \Delta \varphi}$& $ V_0 \exp \left(  \alpha \Delta \varphi \right) \, , \, \alpha =
  \pm 4\sqrt{3}\sqrt{\frac{3}{4}+\mu^2 V_0}$ \\
$ V_0 g^{-n}$ ($n \neq 2$) & $\left[\frac{e^{\eta \Delta \varphi}-4\mu^2V_0 e^{-\eta \Delta \varphi}}{2\sqrt{3}}\right]^{2/n} $&  
$\left[\frac{e^{\eta \Delta \varphi}-4\mu^2V_0 e^{-\eta \Delta \varphi}}{2\sqrt{3}}\right]^{2(2-n)/n}, \quad
\eta=\frac{3n}{2} $ \\
$ \ln g$ & $e^{u(\varphi)} $& $ u e^{2u} \, , \, u=(\mu \sqrt{3}\Delta \varphi)^2 -
  \frac{3}{4 \mu^2}$ \\
$ (\ln g)^2$ & $e^{r(\varphi)} $ & $ r^2 e^{2 r} \, , \, r=\frac{1}{2}\left[e^{u(\varphi)}-\frac{3}{4\mu^2}e^{-u(\varphi)} \right],
\,\, u=2\sqrt{3}\mu\Delta \varphi $ \\
  \hline
\end{tabular}
\end{table}
\end{center}
\end{enumerate}
In this way, the superpotential $\rm S(\Omega,\varphi)$ is known.

For solve (\ref{wdwho}) we assume  
 \begin{equation}
W=e^{\left[ z(\Omega)+\omega(\varphi) \right]},
\end{equation}
we arrive to a set of ordinary differential equations for the functions $\rm z(\Omega)$ and $\omega(\varphi)$
\begin{eqnarray}
\rm 2\frac{dz}{d\Omega}-Q&=&k,\rm \qquad \to \qquad z(\Omega)=\frac{Q+k}{2} \Omega \\
\rm \frac{d^2g}{d\varphi^2} +2\frac{dg}{d\varphi} \frac {d\omega}{d\varphi}&=& 3(k+3)g,\qquad \to \qquad 
\omega(\varphi)=\frac{3k}{2}\int \frac{d\varphi}{\partial_\varphi {(} \ln g {)}} 
- 3\mu^2\int \frac{d{[}{\cal V}(\varphi,\lambda_{eff}){]}}{{(}\partial_\varphi g{)}^2}.
\end{eqnarray}
Thus, the explicit form for the function W becomes
\begin{equation}
W = \exp \left\{ \frac{3k}{2} \left[  \frac{\Omega}{3} +\int
  \frac{d \varphi}{\partial_{\varphi} (\ln g)} \right] +
  \frac{Q}{2} \Omega -  3\mu^2 \int
  \frac{d[{\cal V}(\varphi,\lambda_{eff})]}{\left( \partial_{\varphi} g \right)^2}
  \right\} \, . \label{separation1}
\end{equation}
The constraint (\ref{cons}) can be written as
\begin{equation}
 \rm \partial^2_{\varphi} \omega + \left(\partial_{\varphi} \omega \right)^2 -
\frac{k^2-Q^2}{4} = 0 \, , \label{wdwho2} 
\end{equation}
and
\begin{equation}
\rm  \partial_{\varphi} \omega = \frac{3k}{2\partial_{\varphi} (\ln g)}
 -3\mu^2 \frac{\partial_{\varphi} \left[{\cal V}(\varphi,\lambda_{eff})
 \right] }{\left( 
   \partial_{\varphi} g \right)^2} \, . \nonumber
\end{equation}

Taking into account table I, 
we present the corresponding wave function in each case, in table II.
\begin{center}
\begin{table}[h]
\caption{ \label{wavefunction} Wave function corresponding to  table I}
\begin{tabular}{|c|c|}
\hline
 $\rm g(\varphi)$ &\rm wave function $\Psi$ \\ \hline
 $\rm Exp\left[\pm 3 \Delta \varphi\right]$&  $ \rm \exp \left\{a_2(\pm i\beta_+ +\beta_-) + 
  \left(\frac{k+Q}{2} \right) \Omega \pm \frac{3k}{2} \Delta \varphi\right\}
\,\,e^{  -\frac{e^{3\Omega \pm 3\Delta \varphi}}{\mu}} $  \\ \hline
$\rm {\sqrt{\frac{4\mu^2 V_0}{3}}} sinh(\pm 3 \Delta \varphi)$& $ \rm cosh^{\frac{k}{2}}(\pm 3 \Delta \varphi)\, \,
\exp \left\{a_2(\pm i\beta_+ +\beta_-) +   \left(\frac{k+Q}{2} \right) \Omega  \right\}
\,\,e^{  -\frac{e^{3\Omega} {\sqrt{\frac{4\mu^2 V_0}{3}}} sinh(\pm 3 \Delta \varphi)}{\mu}} $  \\\hline 
 $\rm e^{\frac{\alpha}{2} \Delta \varphi}$& $ \rm \exp \left\{a_2(\pm i\beta_+ +\beta_-) + 
  \left(\frac{k+Q}{2} \right) \Omega + \left(\frac{3k-12\mu^2}{\alpha}\right) \Delta \varphi\right\}
\,\,e^{  -\frac{e^{3\Omega+\frac{\alpha}{2}\Delta \varphi}}{\mu}} $  \\\hline
 $\rm \left[\frac{e^{\eta \Delta \varphi}-4\mu^2V_0 e^{-\eta \Delta \varphi}}{2\sqrt{3}}\right]^{2/n}$ &
  $ \rm \exp \left\{a_2(\pm i\beta_+ +\beta_-) + 
  \left(\frac{k+Q}{2} \right) \Omega + \omega(\varphi)\right\} \,\,e^{  -\frac{e^{3\Omega+g(\varphi)}}{\mu}} $\\
  &$\rm \omega(\varphi)= \frac{k}{2}\left[-\Delta \varphi +\frac{2}{3n}\, Ln\left(4\mu^2 V_0 + e^{2\eta \Delta \varphi} \right)\right]
  -\frac{\mu^2(2-n)}{3nV_0}\, Arctanh\left(\frac{1}{4\mu^2 V_0}e^{2\eta \Delta \varphi} \right)$
 \\\hline
$\rm e^{u(\varphi)} $ &  $ \rm \exp \left\{a_2(\pm i\beta_+ +\beta_-) + 
  \left(\frac{k+Q}{2} \right) \Omega+\frac{k+3-2\mu^2}{4\mu^2}\, Ln\Delta\varphi -\frac{3\mu^2}{2}\Delta \varphi^2\right\}
\,\,e^{  -\frac{e^{3\Omega+u(\varphi)}}{\mu}} $ \\\hline
$\rm e^{r(\varphi)} $ &  $ \rm \exp \left\{a_2(\pm i\beta_+ +\beta_-) + 
  \left(\frac{k+Q}{2} \right) \Omega + \omega(\varphi)\right\}
\,\,e^{  -\frac{e^{3\Omega+r(\varphi)}}{\mu}} $ \\ 
& $\rm \omega(\varphi)= \frac{\sqrt{3}k}{6\mu} \, Arctan\left(\frac{2\mu}{\sqrt{3}}e^{u(\varphi)} \right)
+ 6\mu^3 \Delta \varphi -\frac{\mu}{2}Ln\left(\frac{3}{4\mu^2}+e^{2u(\varphi)}\right)$\\
 &$\rm \qquad +\frac{\mu}{4}  \left[\frac{3}{4\mu^2}e^{u(\varphi)}- e^{-u(\varphi)} \right]
 -\frac{\sqrt{3}}{8} Arctan \left( \frac{\sqrt{3}}{2\mu}e^{-u(\varphi)}\right)$\\ \hline
\end{tabular}
\end{table}
\end{center}

In special case, the third line in  table II, have a damping term that correspond 
to $\rm e^{-S}$ and a plane wave type, 
similar to equation (\ref{wave1}) obtained in the first step. The first line  has 
this behaviour, but corresponds to null scalar potential.

In this way, using the quantum formalism in the sector of inflationary scenario, we found 
that the scalar potential becomes an exponential behaviour, however the
coupling constant is undetermined. The question is, how can we fix the coupling constant?. 
The answer could be in the 
supersymmetric quantum cosmology using differential operators to the Grassmann variables, 
where we present the formalism in  next section.
\section{Supersymmetric quantum mechanics}
In the following we shall apply the supersymmetric quantum formalism at the quantum structure 
obtained in the previous section, to obtain a closed value to the parameters that optimize 
the inflation scenario, i.e, we  do an analysis to the family
obtained in  table I for the function $\rm g(\varphi)$; or in other words, which is the 
constraint on the superpotential function that appears in the quantum level,
 equation (\ref{superpo}), based in  tables I or II?. In this order of ideas,
 we found one integrability condition on the $\rm g(\varphi)$ function, which fix the coupling parameter of our
 problem.
 
For obtain these results, in this section we consider only a reduced
 supersymetry in two bosonic variables $(\Omega,\varphi)$, without consider the anisotropic parameters,
 due that the full problem  do not contemplate the initial condition on our original problem. For instance
 the decomposition of the wave function in the full expansion have 16 components, or  $16\times 16$ matrix
 components, and the solution is very complicated. 
 In this sense, for solve our problem we use reduced bosonic hamiltonian, eqn (\ref{wdw2}), and in consequence, 
 one reduced supersymmetry.

The idea of Witten \cite{witten} is to find the supersymmetric supercharges operators
$\rm Q$, $\rm \bar Q$ that produce a superhamiltonial ${\cal H}_{susy}$,  that
satisfies the closed superalgebra
\begin{equation}
\rm {\cal H}_{susy}=\frac{1}{2}\left[ Q, \overline Q \right], \qquad \left[\overline Q, \overline Q \right]= \left[ Q, Q \right]=0, 
\end{equation}
where the superhamiltonian ${\cal H}_{susy}$ has the following form
\begin{equation}
\rm {\cal H}_{susy}={\cal H}_b + \frac{\partial^2 S}{\partial q^\mu \partial q^\nu} \left[ \overline \psi^\mu, \psi^\nu\right], \label{hsusy}
\end{equation}
with ${\cal H}_b$ is the bosonic hamiltonian (\ref{wdw2}) taking the constant c=0, then the supersymmetric approach  will only be
applied to reduced hamiltonian, and S is the corresponding superpotential function that is related with the potential term that 
appear in the bosonic hamiltonian, i.e,
 have the same structure that in the quantum level, proposed in the last section. This idea was applied in reference \cite{socorro} for
 all Bianchi Class A models without matter content, and in \cite{socorro1} to FRW cosmological model. For example in reference \cite{moniz}
it is explained that in particular, we can formulate a 
 {\it particle dynamics} in a potential $\rm V(q^\mu)$ on a curved manifold and supersymmetry requires that the potential $\rm V(q^\mu)$
 is derivable from a globally defined superpotential $\rm S(q^\mu)$ 
 via $\rm V(q^\mu)=\frac{1}{2}G^{\mu\nu}(q) \frac{\partial S(q)}{\partial q^\mu} \frac{\partial S(q)}{\partial q^\nu}$, where 
 $\rm G^{\mu\nu}(q^\mu)$ is the metric in the curved space. This equation is represented in the quantum level by eq. (\ref{hj}). 

In this approach, a supersymmetric state with $\rm Q|\psi>=0$ is automatically a zero energy ground state,
in  a similar way that in the quantum regime. This simplifies the problem of finding supersymmetric ground
state because the energy is known as priori and the factorization of
${\cal H}_{susy}|\psi>=0$ into $\rm Q|\psi>=0$, $\rm \bar Q|\psi>=0$ often provides a simpler first 
order equation for the ground state wavefunction. The simplicity of this factorization is related to the
solubility of certain bosonic hamiltonians. In this work, as in others, we find for the
empty (+) and filled (-) sector of the expantion of the wavefunction in this approach, 
in the sector of the fermion Fock space zero energy solutions $\rm |{\cal A}_\pm>=e^{\pm S}|\pm>$
where ${\cal A}_\pm$ are the corresponding components for the empty and filled fermionic sector.

The corresponding supercharges that satisfy the superalgebra, when we consider the bosonic hamiltonian 
given by equation (\ref{wdw2}) become
\begin{eqnarray}
\rm Q&=&\rm  \psi^\mu \left[ \frac{\partial}{\partial q^\mu} + \frac{\partial S}{\partial q^\mu} \right], \nonumber\\
\rm \overline Q&=& \rm  \overline \psi^\nu \left[\frac{\partial}{\partial q^\nu} - \frac{\partial S}{\partial q^\nu} \right]. \label{barq}
\end{eqnarray}
We consider the following algebra for the fermionic variables \cite{socorro}
\begin{equation}
\rm  \left\{\psi^\mu,\overline\psi^\nu\right\}=\eta^{\mu\nu},\qquad
\qquad \left\{\overline\psi^\mu,\overline\psi^\nu\right\}=\left\{\psi^\mu,\psi^\nu\right\}=0,
\end{equation}
and the corresponding representation
\begin{equation}
\rm \overline \psi^\nu=\theta^\nu, \qquad \psi^\mu=\eta^{\mu \nu} \frac{\partial}{\partial \theta^\nu}.
\end{equation}

Equation (\ref{barq}) are
\begin{eqnarray}
\rm Q&=&\rm -  \left[ \frac{\partial}{\partial q^0} + \frac{\partial S}{\partial q^0} \right] \frac{\partial}{\partial \theta^0}
+ \left[ \frac{\partial}{\partial q^1} + \frac{\partial S}{\partial q^1} \right] \frac{\partial}{\partial \theta^1}, \nonumber\\
\rm \overline Q&=& \rm  \theta^0 \left[\frac{\partial}{\partial q^0} - \frac{\partial S}{\partial q^0} \right]
+\theta^1 \left[\frac{\partial}{\partial q^1} - \frac{\partial S}{\partial q^1} \right]. \label{susy}
\end{eqnarray}
The decomposition of the wavefuntion becomes
\begin{equation}
\rm \Xi(\Omega, \varphi)= {\cal A}_+ + {\cal B}_0 \theta^0+ {\cal B}_1 \theta^1+{\cal A}_- \theta^0 \theta^1,
\end{equation}
where the coordinates fields are $\rm q^\mu=(q^0,q^1)=(\Omega,\varphi)$,  ${\cal A}_\pm$, $\rm {\cal B}_0, {\cal B}_1$ are the bosonic
and fermionic contributions to the wavefunction.

The supersymmetric equations $\rm Q|\Xi>=0$, $\rm \bar Q|\Xi>=0$ are 
\begin{eqnarray}
\rm Q \Xi&=&\rm -  \left[ \frac{\partial}{\partial q^0} + \frac{\partial S}{\partial q^0} \right] \frac{\partial}{\partial \theta^0}
\left[{\cal A}_+ + {\cal B}_0 \theta^0+ {\cal B}_1 \theta^1+{\cal A}_- \theta^0 \theta^1 \right]\nonumber\\
&&+ \left[ \frac{\partial}{\partial q^1} + \frac{\partial S}{\partial q^1} \right] \frac{\partial}{\partial \theta^1}
\left[{\cal A}_+ + {\cal B}_0 \theta^0+ {\cal B}_1 \theta^1+{\cal A}_- \theta^0 \theta^1 \right]
, \label{no-bar}\\
\rm \overline Q \Xi&=& \rm  \theta^0 \left[\frac{\partial}{\partial q^0} - \frac{\partial S}{\partial q^0} \right]
\left[{\cal A}_+ + {\cal B}_0 \theta^0+ {\cal B}_1 \theta^1+{\cal A}_- \theta^0 \theta^1 \right] \nonumber\\
&&+\theta^1 \left[\frac{\partial}{\partial q^1} - \frac{\partial S}{\partial q^1} \right]
\left[{\cal A}_+ + {\cal B}_0 \theta^0+ {\cal B}_1 \theta^1+{\cal A}_- \theta^0 \theta^1 \right].
 \label{bar}
\end{eqnarray}

Then  (\ref{bar})  gives the following set of differential equations
\begin{eqnarray}
\rm \theta^0: && \rm  \left[\frac{\partial {\cal A}_+}{\partial q^0} - {\cal A}_+\frac{\partial S}{\partial q^0} \right]=0, \label{a+}\\
\rm \theta^1: && \rm \left[\frac{\partial {\cal A}_+}{\partial q^1} - {\cal A}_+\frac{\partial S}{\partial q^1} \right]=0, \label{a++}\\
\rm \theta^0\theta^1: &&\rm  \left[\frac{\partial {\cal B}_1}{\partial q^0} - {\cal B}_1\frac{\partial S}{\partial q^0} \right]-
\left[\frac{\partial {\cal B}_0}{\partial q^1} - {\cal B}_0\frac{\partial S}{\partial q^1} \right]=0,\label{b1}
\end{eqnarray}
whose solutions to equations (\ref{a+},\ref{a++}) are
\begin{equation}
\rm {\cal A}_+=a_+\, e^{S}, 
\end{equation}
On the other hand, eq. (\ref{no-bar}) gives
\begin{eqnarray}
\mbox{free term}:&& -\left[ \frac{\partial {\cal B}_0}{\partial q^0} + {\cal B}_0\frac{\partial S}{\partial q^0} \right]
+ \left[ \frac{\partial {\cal B}_1}{\partial q^1} + {\cal B}_1\frac{\partial S}{\partial q^1} \right] =0, \label{b01}\\
\theta^1: && \left[ \frac{\partial {\cal A}_-}{\partial q^0} + {\cal A}_-\frac{\partial S}{\partial q^0} \right]=0, \label{a-}\\
\theta^0: && \left[ \frac{\partial {\cal A}_-}{\partial q^1} + {\cal A}_-\frac{\partial S}{\partial q^1} \right]=0, \label{a--}
\end{eqnarray}
where (\ref{b01}) can be written as
\begin{equation}
\rm \eta^{\mu\nu} \left( \partial_\mu {\cal B}_\nu +{\cal B}_\nu \partial_\mu S \right) =0, \label{bmu1}
\end{equation}
considering the following ansatz for the fields ${\cal B}_\mu$ 
\begin{equation}
{\cal B}_\nu=e^{S}\, \partial_\nu f_+,
\end{equation}
 (\ref{b1}) is satisfied identically, and  (\ref{bmu1}) is
\begin{equation}
\rm \eta^{\mu\nu} \left( \partial_\mu \partial_\nu f_++\partial_\nu f_+ \partial_\mu S +\partial_\nu f_+ \partial_\mu S \right) =
\eta^{\mu\nu} \left( \partial_\mu \partial_\nu f_++2\partial_\nu f_+ \partial_\mu S\right) =0, \label{bmu2}
\end{equation}
where a possible solution is $\rm f_+=h(\Omega \pm \varphi)$, and h is  any function dependind to the argument,
given the following constrain on the superpotential function 
\begin{equation}
 \frac{\partial S}{\partial \Omega}=\pm \frac{\partial S}{\partial \varphi}, \label{constraint}  
\end{equation}
and considering the structure of the superpotencial (\ref{superpo}) we find one condition on integrability
over the function $\rm g(\varphi)$, given 
\begin{equation} 
\rm g(\varphi)=g_0 e^{\pm 3 \Delta \varphi}.
\end{equation}
and taking into account tables I or II, we
obtain the following constraint in the parameter $\alpha$ of the models when this last equation (\ref{constraint}) 
is satisfied 
$$\frac{\alpha}{2}=\pm 3,$$
so, only the exponential scalar potential can survive in  table I, and the coupling constant
become $\alpha=\pm 6$, given the scalar potencial $\rm V(\phi)=V_0 e^{-\sqrt{3}\Delta\phi}$. 
In this way, supersymmetric quantum mechanics fix the values for the $\alpha$ parameter, being valid
the argument introduced in the quantum scheme.

Equations, (\ref{a-}, \ref{a--}) can be written as
\begin{equation}
\rm \frac{\partial {\cal A}_-}{\partial q^\mu}+ {\cal A}_- \frac{\partial S}{\partial q^\mu}=0, \qquad 
\frac{1}{{\cal A}_-} \frac{\partial {\cal A}_-}{\partial q^\mu}=- \frac{\partial S}{\partial q^\mu} \qquad \rightarrow
\qquad \frac{\partial Ln{\cal A}_-}{\partial q^\mu }=-\frac{\partial S}{\partial q^\mu},
\end{equation}
with solution 
\begin{equation}
\rm {\cal A}_-=a_- e^{-S}, \label{solucion:a-}
\end{equation}
then, the set of contributions for the supersymmetric wave functions are find,
\begin{eqnarray*}
{\cal A}_\pm &=&\rm a_\pm e^{\pm S}\\
{\cal B}_0 &=& e^S\partial_0 (f_+)\\
{\cal B}_1 &=& e^S\partial_1 (f_+)
\end{eqnarray*}
It is interesting to note that supersymmetry is very restrictive because exist more constraints equations applied to the wave function. In this
sense, we observe a tendency for supersymmetric vacua to remain close to their
semi-classical limits, because the exact solutions found are also the lowest-order WKB approximation.
\section{Conclusions}
Using the quantum formalism in the  inflationary scenario, we find that the scalar potential has an  exponential behaviour 
as a good candidate. However, the
coupling constant is undetermined. The question was, how  can we  fix the value of the  coupling constant? The answer was in the 
supersymmetric quantum cosmology using differential operators to the Grassmann variables, where
the coupling constant is found under one condition of integrability
on the function $\rm g(\varphi)=g_0 e^{\pm 3 \Delta \varphi}$, and taking into account tables I or II, $\frac{\alpha}{2}=\pm 3$.
So, the main goal in this paper was to fix the value for the coupling constant to the inflationary scenario $\lambda=\frac{\alpha}{2}=\pm 3$ using the 
supersymmetric approach, when the quantum approach only gives the general structure for the scalar potential. Also we find exact solutions
in both regimes. In the quantum level, we found that the possible solutions become the contributions to the
empty(+) and filled (-) sector of decomposition to the wavefunction in the supersymmetric approach. 

\acknowledgments{We thank T. Matos, C. Escamilla-Rivera, N. L\'opez, M. Reyes, 
L. Ure\~na and M. Sabido for critical comments of this manuscript. 
This work was supported in part by  DINPO(2009) and
Promep grant UGTO-CA-3.
Many calculations where done by Symbolic Program REDUCE 3.8. 
This work is part of the collaboration within the Instituto Avanzado de
Cosmolog\'{\i}a. The authors also thank the anonymous referee for his fruitful comments.}


\begin{thebibliography}{99}


\bibitem{wssa} W. Guzm\'an, M. Sabido, J. Socorro and L. Arturo Ure\~na-L\'opez, 
	Int. J. Mod.  Phys. D {\bf 16} (4), 641-653 (2007), [gr-qc/0506041]	
\bibitem{celia} C. Escamilla-Rivera, {\it Lo que la supersimetr\'{\i}a local puede hacer por la cosmolog\'{\i}a}, 
	Master thesis, DCeI, Universidad de Guanajuato (2019).
\bibitem{scalar-1} P. J. E. Peebles and B. Ratra, The cosmological constant and dark energy, Rev. Mod.
Phys. {\bf 75}, 559 (2003).
\bibitem{scalar-2} D. Lyth and A. Liddle, Cosmological Inflation and Large Scale Structure (Cambridge
University Press, 2000).
\bibitem{scalar-3}  L. Arturo Urena-Lopez and T. Matos, A new cosmological tracker solution for
quintessence, Phys. Rev. D {\bf 62}, 081302 (2000).
\bibitem{sca4} V. Sahni and L.-M.Wang, A new cosmological model of quintessence and dark matter,
Phys. Rev. D {\bf 62}, 103517 (2000).
\bibitem{sca5} T. Matos and L. A. Urena-Lopez, On the nature of dark matter, Int. J. Mod. Phys.
D {\bf 13}, 2287 (2004).
\bibitem{sca6} A. Arbey, J. Lesgourgues and P. Salati. Quintessential haloes around galaxies, Phys.
Rev. D {\bf 64}, 123528 (2001).
\bibitem{sca7} U. Alam, V. Sahni and A. A. Starobinsky, The case for dynamical dark energy revisited,
J. Cosmol. Astropart. Phys. {\bf 0406}, 008 (2004).
\bibitem{sca8} V. Sahni, Lect. Notes Phys. {\bf 653}, 141 (2004).
\bibitem{sca9} M. Tegmark, What does inflation really predict? J. Cosmol. Astroport. Phys. {\bf 0504},
001 (2005).
\bibitem{copeland1} E.J. Copeland, A.R. Liddle and D. Wands, Phys. Rev. D {\bf 57}, 4686 (1998) (and references therein).
\bibitem{tsuji} S. Tsujikawa, Phys.Rev. D  {\bf 62}, 043512  (2000),  [arXiv:hep-ph/0004088].
\bibitem{copeland} E.J. Copeland, M. Sami and S. Tsujikawa, {\it Dynamics od dark energy}, [arXiv:hep-th/0603057]
\bibitem{bene} J. Bene and R. Graham, Phys. Rev. D {\bf 49}, 799 /1994); R. Graham, Phys. Rev. Lett.
{\bf 67}, 1381 (1991).
\bibitem{osb} O. Obreg\'on, J. Socorro and J. Ben\'{\i}nez, Phys. Rev. D {\bf 47}, 4471 (1993).
\bibitem{lidsey} J.E. Lidsey, Phys. Rev. D {\bf 52}, R5407 (1995).
\bibitem{socorro} J. Socorro and E.R. Medina, Phys. Rev. D {\bf 61}, 087702-1 (2000).
\bibitem{taquion} A. Sen, Mod. Phys. Lett. A {\bf 17}, 1797 (2002); Mod. Phys. Lett. A {\bf 18}, 4869 (2003);
	 V. Gorini, A.Y. Kamenshchik, U. Moschella and V. Pasquier, Phys. Rev. D {\bf 69},
	123512 (2004); H. Garcia-Compean, G. Garcia-Jimenez, O. Obregon and C. Ramirez, Phys. Rev. D
		{\bf 71}, 063517 (2005).
\bibitem{bohm} D. Bohm, Phys. Rev. {\bf 85}, 166, (1952).
\bibitem{OS} O. Obreg\'on and J. Socorro,  Int. J. Theor. Phys. 	{\bf 35}, 1381 (1996).
\bibitem{HaHa} J. Hartle and S.W. Hawking, Phys. Rev. D {\bf 28}, 2960 (1983).
\bibitem{hawking} S.W. Hawking, Nucl. Phys. B {\bf 239}, 257 (1984). 
\bibitem{gibbons} G.W. Gibbons and L.P. Grishchuk, Nucl. Phys. B {\bf 313}, 736 (1989).
\bibitem{ruffini} {\it Quantum cosmology}, Edited by Li Zhi Fang and Remo Ruffini, Advances Series 
in Astrophysics and Cosmology Vol. 3, World Scientific (1987) 
\bibitem{ryan} M.P. Ryan and L.C. Shepley, {\it Homogeneous Relativistic Cosmologies},
        Princeton University Press, Princeton, New Jersey (1975).
\bibitem{kasper} U. Kasper, Class. Quantum Grav. {\bf 10}, 869 (1993).
\bibitem{cimento} L.P. Chimento and A.S. Jakubi, Int. J. Mod. Phys. D {\bf 5}, 71 (1996).
\bibitem{witten} E. Witten, Nucl. Phys. B {\bf 188}, 513 (1981).
\bibitem{socorro1} J. Socorro, Rev. Mex. F\'is. {\bf 48} (2) 112-117 (2002).
\bibitem{moniz} P.V. Moniz, Gen. Rel. and Grav. {\bf 38}, 577-592 (2006).

		
\end{thebibliography}
\end{document}